# The Problem of Civil and Criminal Identification - Bayesian Networks Approach

**MARINA ANDRADE**

marina.andrade@iscte-iul.pt

**MANUEL ALBERTO M. FERREIRA**

manuel.ferreira@iscte-iul.pt

## ABSTRACT

DNA evidence use in problems of civil and criminal identification is becoming greater. The necessity of evaluating the weight of that evidence may be accomplished using one of the most known powerful tools: the Bayesian networks. In the current paper this will be illustrated through the presentation of a civil identification problem and of a criminal identification problem.

## 1. INTRODUCTION

It is intended to exemplify how to apply Bayesian networks in civil and criminal identification problems.

The first objective is to give a methodology, and the appropriate tools, to use correctly a DNA profiles database in the problem of civil identification when there is a partial match between the genetic characteristic of an individual whose body was found, one volunteer who claimed a family member disappearance and one sample belonging to the DNA database.

In section 2 the civil identification case to be studied is presented and discussed. The Bayesian network, that allows the efficient probabilities computation, determinant to evaluate the hypothesis in comparison, is presented. In section 3, real examples clarifying the application are exhibited. And in section 4 a brief discussion is outlined related with this objective.

The second objective is to illustrate the use of biological information in crime scene identification problems, an example of a criminal identification problem.

In section 5 a crime scene, the correspondent evidence, *E*, and the hypotheses to be considered are presented. In section 6 the Bayesian network built expressly to perform the calculations is shown. And in section 7 the numerical results will be presented and preliminarily discussed. Finally in section 8 a brief discussion, related with this second objective is presented.

General conclusions and references are presented at the end of this paper.

## 2. CIVIL IDENTIFICATION PROBLEM

Frequent examples of civil identification problems are the case of a body identification, jointly with information of a missing person belonging to a known family, or the identification of more than one body resultant of a disaster or an attempt. And even immigration cases in which it is important to establish family relations.

The establishment and use of DNA database files for a great number of European countries was an incentive to study the mentioned problems and the use of these database files for identification, see (1-4). In this context it may be useful when unidentified corpses appear and may be identified by comparison of their DNA profiles with possibly family volunteer's profiles.

The Portuguese law nº 5/2008 establishes the principles for creating and maintaining a database of DNA profiles for identification purposes[1], and regulates the collection, processing and conservation of samples of human cells, their analysis and collection of DNA profiles, the methodology for comparison of DNA profiles taken from the samples, and the processing and storage of information in a computer file[2].

So the database is, in general terms, composed of a file containing information of samples from convicted offenders with 3 years of imprisonment or more - $\alpha$; a file containing the information of samples of volunteers -$\beta$; a file containing information on the "problem samples" or "reference samples" from corpses, or parts of corpses, or things in places where the authorities collect samples - $\gamma$.

It matters to study civil identification problems, mainly if there is a partial match between the genetic characteristic of an individual whose body was found and one volunteer who claimed a family member disappearance and one sample in the database file $\gamma$.

### 2.1 PARTIAL MATCH

When there is an individual claiming for a disappeared person who gives his/her genetic characteristic, $C_{Vol}$[3], to be compared with the genetic characteristic of a body found, the first action to do is to check if there is a match between the genetic characteristic of the individual whose body was found, $C_{BF}$, and any sample of the DNA file, $\gamma$ - sample, which is named "problem samples".

If there is a partial match between the genetic profile of the individual whose body was found and one sample in the file $\gamma$, the evidence now is $E = (C_{BF}, \gamma - sample, C_{Vol})$.

Then it follows the establishment of the hypotheses of interest: the identification hypothesis ($H_{ID}$) versus the nonidentification hypothesis ($H_{not\ ID}$):

$H_{ID}$: It is possible to reach an identification of the individual whose body was found.

*vs.*

---

[1] This project is now stopped due budget constraints.
[2] The implementation of this process is not going very well. In the moment the number of samples in database is not significant.
[3] That, of course, integrates the $\beta$ file.

$H_{not\ ID}$: It is not possible to reach an identification of the individual whose body was found.

After checking the possibility of a partial match between the profile of the individual whose body was found, $C_{BF}$, the sample in the file $\gamma$, $\gamma$ - sample, and the volunteer, $C_{Vol}$, two different comparisons are made in order to obtain a measure either of the possible genetic relation between the individual whose body was found with the $\gamma$ - sample (*bf_match_gs?*), or of the possible genetic relation between the individual whose body was found and the volunteer (*bf_match_vol?*). The possible answers are yes or no.

So, the resulting states are:

- *A* (yes, no) - defines the possibility of genetic relationship between the individual whose body was found and the $\gamma$ - sample but not the volunteer.

- *B* (no, yes) - defines the possibility of genetic relationship between the individual whose body was found and the volunteer but not the $\gamma$ - sample.

- *C* (yes, yes) - defines the possibility of genetic relationship between the individual whose body was found and both the volunteer and the $\gamma$ - sample.

- *D* (no, no) - defines the possibility of genetic relationship between the individual whose body was found neither with the volunteer nor with the $\gamma$ - sample.

States *A*, *B* define the identification hypothesis, $H_{ID}$, and *C*, *D* define the non-identification hypothesis, $H_{not\ ID}$. State *B* is a particular case: the simple problem studied in (3). Each one of the four possible states probabilities provides a measure for each event, and the four are pairwise incompatible.

Following the probabilities computation it is important to compare state *D* versus *A*, *B*, *C*, i.e., to evaluate the event "the individual whose body was found is not genetically related either with the $\gamma$ - sample or the volunteer". With this comparison it is intended to evaluate the situation "*the genetic information of the individual whose body was found is not compatible with the other genetic information available*" and "*the genetic information of the individual whose body was found is compatible with at least one of the remaining genetic types of information*".

If *D* is accepted the process ends. And the body genetic information joins the file $\gamma$ in the database. If *D* is discarded then it is necessary to perform a comparison between *A*, *B* and *C* events. If *C* is accepted the process ends and police intelligence investigations must be done. If *C* is discarded, finally *A* and *B* must be compared. If *A* is accepted the individual whose body was found is related with the $\gamma$ - sample. If *B* is accepted the conclusion is that the individual whose body was found is a volunteer relative.

## 3. BAYESIAN NETWORK FOR THE CIVIL IDENTIFICATION PROBLEM

The comparisons described above are performed through the respective probabilities' events ratios: the likelihood ratios, (5). The hypothesis with the greatest probability is the accepted one. Thus, the probabilities associated to the states A, B, C and D must be computed. To do so a lot of intermediary conditional probabilities computation, that are impossible to do with algebraic manipulations, must be done.

To overcome this situation those probabilities will be computed using the Bayesian network, see (6-7), in the Fig. 1[4].

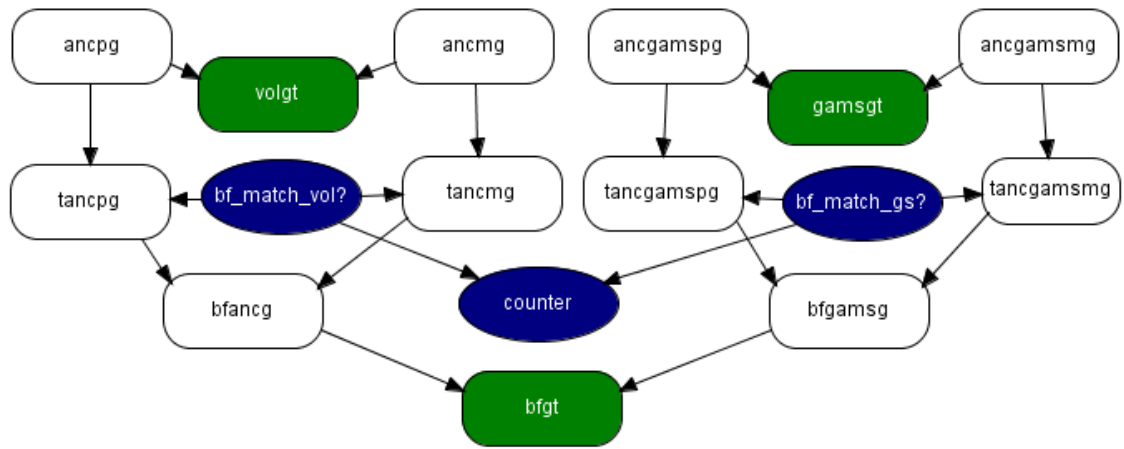

**Fig. 1.** Network for civil identification with one volunteer and one $\gamma - sample$

The nodes **ancpg**, **ancmg**, **ancgampg,** and **ancgammg** are of class founder: a network with only one node which states are the alleles in the problem where the respective frequencies in the population are specified and represent the volunteer's ancient paternal and maternal inheritance.

The nodes **volgt**, **gamsgt,** and **bfgt** are of class genotype: the volunteer, the $\gamma$ - sample and the body found genotypes.

Nodes **tancmg**, **tancpg**, **tancgamspg,** and **tancgamsmg** specify whether the correspondent allele is or is not the same as the volunteer and the same as the $\gamma$ - sample.

If *bf_match_vol?* is true then the volunteer's allele will be identical with the body found allele, otherwise the allele is randomly chosen in the population and if *bf_match_gs?* is true then the $\gamma$ - sample's allele will be identical with the body found allele, otherwise the allele is randomly chosen in the population.

The nodes **bfancg,** and **bfgamsg** define the Mendel inheritance in which the allele of the individual whose body was found is chosen at random from the ancient's paternal and maternal gene.

Node counter counts the number of true states of the preceding nodes, accounting the results for the *A*, *B*, *C*, *D* possible events.

---

[4]The networks mentioned in this work were implemented using Hugin software: www.hugin.com

## 3.1 EXAMPLES

To exemplify the described methodology, in Table 1 the allele frequencies, real ones, for some genetic markers and, for each marker, possible evidence profiles for the body found $C_{BF}$, the $\gamma$ - sample and the volunteer $C_{Vol}$ are presented.

**Table 1.** Allele frequencies and genetic profiles.

| Marker | Allele Frequencies | | | | $\{C_{BF}, \gamma - sample, C_{Vol}\}$ |
|---|---|---|---|---|---|
| D21S11 | $p_{28}$ 0.1647 | $p_{29}$ 0.2136 | $p_{30}$ 0.2437 | $p_{31.2}$ 0.1138 | $\{(29,30), (28,30), (29,31.2)\}$ |
| F13A1 | $p_5$ 0.1985 | $p_6$ 0.2890 | $p_7$ 0.3377 | $p_8$ 0.0112 | $\{(6,7), (7,8), (5,6)\}$ |
| TH01 | $p_6$ 0.2044 | $p_7$ 0.1696 | $p_9$ 0.1984 | $p_{9.3}$ 0.2748 | $\{(7,9), (9,9.3), (6,7)\}$ |
| TPOX | $p_8$ 0.5053 | $p_9$ 0.0974 | $p_{10}$ 0.0647 | $p_{11}$ 0.2893 | $\{(8,11), (8,10), (9,11)\}$ |
| VWA31 | $p_{15}$ 0.1216 | $p_{16}$ 0.2300 | $p_{17}$ 0.2649 | $p_{18}$ 0.1859 | $\{(16,17), (15,17), (16,18)\}$ |

In Table 2 the state probabilities, the node counter states, see Fig. 1, are presented.

**Table 2.** State probabilities.

| States | D21S11 | F13A1 | TH01 | TPOX | VWA31 |
|---|---|---|---|---|---|
| A | 0.5322 | 0.3296 | 0.4987 | 0.2661 | 0.4548 |
| B | 0.1296 | 0.2226 | 0.1978 | 0.2688 | 0.2251 |
| C | 0.2274 | 0.1904 | 0.1692 | 0.1539 | 0.2092 |
| D | 0.1108 | 0.2574 | 0.1343 | 0.3112 | 0.1109 |

And in Table 3 the decisions, consequence of the procedures proposed in section 2.1, are presented for each example evidence profile.

**Table 3.** Decisions for each evidence profile.

| Evidence Profiles | Decision |
|---|---|
| $\{(29,30), (28,30), (29,31.2)\}$ | Police intelligence investigations must be done |
| $\{(6,7), (7,8), (5,6)\}$ | The individual whose body was found is a volunteer relative |
| $\{(7,9), (9,9.3), (6,7)\}$ | Police intelligence investigations must be done |
| $\{(8,11), (8,10), (9,11)\}$ | The individual whose body was found is a volunteer relative |
| $\{(16,17), (15,17), (16,18)\}$ | Police intelligence investigations must be done |

## 4. CIVIL IDENTIFICATION PROBLEM DISCUSSION

Using the Bayesian network built expressly for civil identification problem, in which there is a partial match between an individual whose body was found, a volunteer who claimed a relative disappearance supplying his/her own genetic information and a DNA database file sample existent, it is possible to perform the sequence of three hypothesis tests described above. Thus, it is possible to decide first if an identification is possible or not; second if an effective identification is possible or not; third to make the identification. So, with a procedure technically simple, it is possible to make an adequate and correct use of a DNA database.

As the examples illustrate, the procedure leads almost surely to a decision: whether it is to close the case identifying the individual, or concluding that it is not possible any identification, or to go on with the police investigations.

## 5. CRIME SCENE INVESTIGATION

A crime has been committed. Two persons, $V_1$ and $V_2$, were murdered. One mixture trace was found. $S_1$ and $S_2$ are potential suspects. $S_1$ and $S_2$ DNA profiles were measured and considered to be compatible with the mixture trace.

Being possible that a fight occurred during the assault, producing some material, it is acceptable that the individuals who perpetrated the crime could have left some of their material in the trace.

To analyse the crime scene, in this section, it will be presented the evidence, $E$, and the hypotheses to be considered.

To summarize the evidence, it is presented in Table 4 the DNA profiles of the victims' and the suspect's, $V_1$, $V_2$, $S_1$, $S_2$, and the trace found at the crime scene, $E$.

**Table 4.** Two victim's and two suspect's DNA profiles and evidence.

|       | $V_1$ | $V_2$ | $S_1$ | $S_2$ | $E$ |
|-------|-------|-------|-------|-------|-----|
| **TH01**  | 9,9.3 | 9,9.3 | 7,8   | 6,9   | 6,7,8,9,9.3 |
| **F13A1** | 5,7   | 5,6   | 3.2,5 | 6,7   | 3.2,5,6,7 |
| **FGA**   | 22,26 | 22,23 | 24,24 | 19,24 | 19,22,23, 24,26 |

In Table 5 the allele frequencies, for each marker found in the trace, are presented.

**Table 5.** Allele frequencies.

|       | $p_6$ | $p_7$ | $p_8$ | $p_9$ | $p_{9.3}$ |
|-------|-------|-------|-------|-------|-----------|
| **TH01** | 0.2044 | 0.1696 | 0.1386 | 0.1984 | 0.2748 |

|       | $p_{3.2}$ | $p_5$ | $p_6$ | $p_7$ | |
|-------|-----------|-------|-------|-------|--|
| **F13A1** | 0.0806 | 0.1985 | 0.2890 | 0.3377 | |

|       | $p_{19}$ | $p_{22}$ | $p_{23}$ | $p_{24}$ | $p_{26}$ |
|-------|----------|----------|----------|----------|----------|
| **FGA** | 0.0684 | 0.1740 | 0.1606 | 0.1325 | 0.0321 |

The allele frequencies in Table 5 are the Portuguese population frequencies collected in the database "The Distribution of Human DNA-PCR Polymorphisms", since the mentioned case is supposed to have occurred in Portugal.

The crime trace can contain DNA from up to four unknown contributors, in addition to the victims and/or the suspects.

If the DNA of $S_i$ with $i = 1, 2$ is presented in the trace this will place him/her at the crime scene and consequently as one of the possible perpetrators.

The court has to determine if each suspect is or is not guilty. The hypotheses to be evaluated are:

$H_1$: $S_1$ is a contributor to the trace but $S_2$ is not, given the evidence.

$H_2$: $S_2$ is a contributor to the trace but $S_1$ is not, given the evidence.

$H_3$: $S_1$ and $S_2$ are both contributors to the trace, given the evidence.

$H_4$: Neither $S_1$ nor $S_2$ are contributors to the trace, given the evidence.

The respective events probabilities are denominated $p_{10}$, $p_{02}$, $p_{12}$, $p_{00}$, where 0 mentions the absence of the respective, in order, individual DNA in the trace. So:

If $p_{00} > p_{10} + p_{02} + p_{12}$ the two suspects are acquitted. If not it must be seen if $p_{12} > p_{10} + p_{02}$ case at which the two suspects are both placed at the crime scene. If not $p_{10}$ must be compared with $p_{02}$. If $p_{10} > p_{02}$ the evidence favours the presence of $S_1$ at the crime scene and the acquaintance of $S_2$. The contrary happens when $p_{02} > p_{10}$.

## 6. BAYESIAN NETWORK FOR THE CRIME SCENE INVESTIGATION

The probabilities referred above are very hard to compute algebraically, demanding a great use of Bayes' Law because of the number of the dependencies to be considered. So they will be computed using the Bayesian network of Fig. 2.

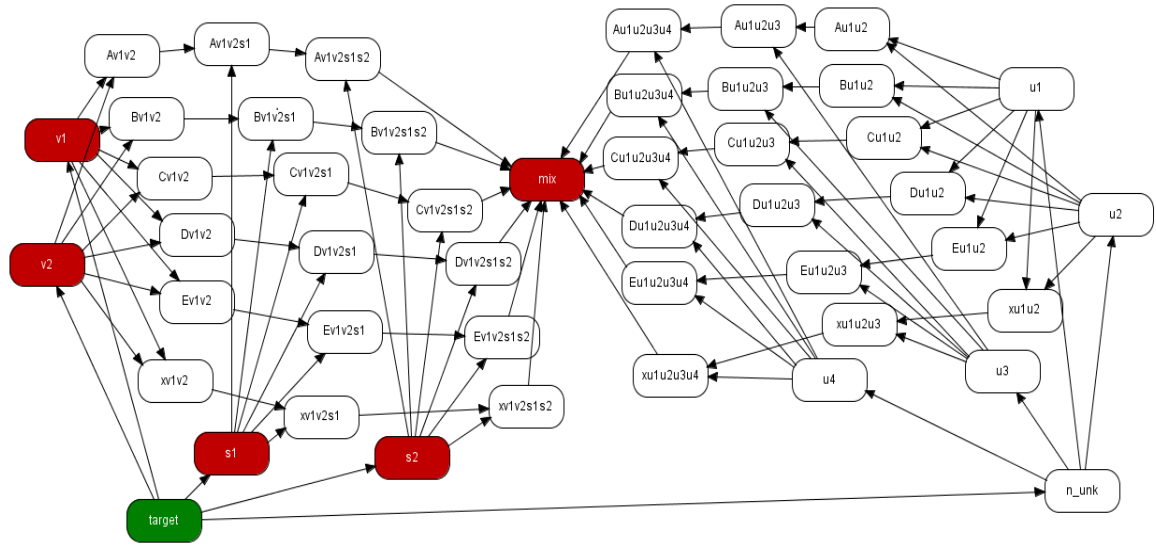

**Fig. 2.** Marker network.

Nodes $v_i$, $i = 1, 2$, $s_j$, $j = 1, 2$ and $u_k$, $k = 1, 2, 3, 4$, in Fig. 2 are themselves Bayesian networks that represent the genetic structure and inheritance of each individual - the victims, the suspects and the unknowns, respectively - and have all the same structure. The $v_i$, $i = 1, 2$ and $s_j$, $j = 1, 2$ are represented in red colour meaning that the respective profiles are known and constitute data of the problem. The nodes in white, below the node mix, that represents the mixture and is also in red colour because it is comprised by known data ($E$), represent the relations in which the nodes in red may contribute to the mixture. The nodes in white, above the node mix, except the $u_k$, $k = 1, 2, 3, 4$ and **n_unk** - that is a counter for the number of unknowns in the mixture – represent the relations in which the $u_k$, $k = 1, 2, 3, 4$ may contribute to the mixture. Node **target**, in green colour, collects the states and the respective probabilities.

As it is mandatory to consider the possible contribution of till four unknown individuals to the mixture, the number of admissible states jumps to 80, numbered from 0 - no one in the mixture - to 79 - the two victims, the two suspects and the four unknowns are all in the mixture. Of course, these two states are unrealistic and there are other ones also unrealistic because are incompatible with the minimum number of contributors to the mixture, according to the evidence inserted. These unrealistic states are discarded by the network but must be considered conceptually in its building.

Among the realistic states only a few ones are interesting to the problem: the corresponding to the hypothesis's events defined above.

## 7. NUMERICAL RESULTS

For marker TH01, alleles 6, 7, 8, 9, 9.3 are considered, Table 4, and so they are represented in the Fig. 1 Bayesian network by A, B, C, D, E, respectively. When considering marker F13A1, the alleles are 3.2, 5, 6, 7, corresponding to A, B, C, D. E is considered with 0 frequency. In marker FGA the alleles are 19, 22, 23, 24, 26 corresponding to A, B, C, D, E. In any case x accumulates the remaining frequencies of the non-considered alleles for each marker.

The results obtained using Table 4 data together with Table 5 frequencies are in Table 6, where the values in line rescale are constituted by the ratios of the products of the values in the respective column[5] by the total sum of the four products. The values in this line are the used ones in the tests described in section 5.

**Table 6.** Results.

|  | $p_{00}$ | $p_{12}$ | $p_{10}$ | $p_{02}$ |
|---|---|---|---|---|
| **TH01** | 0.0830 | 0.5029 | 0.2773 | 0.1367 |
| **F13A1** | 0.0986 | 0.4544 | 0.3279 | 0.1187 |
| **FGA** | 0.0378 | 0.4398 | 0.0820 | 0.4398 |
| **Rescale** | 0.0027 | 0.8709 | 0.0646 | 0.0618 |

Following the procedure outlined in section 5 the conclusion is that both suspects are placed at the crime scene – note the great value of $p_{12} = 0.8709$. For TH01 and F13A1, alone, the conclusion is the same. But for FGA this does not happens. Note that, $p_{12} = p_{02}$. This is justified by the fact that in marker FGA there are two rare alleles, $p_{19} = 0.0684$ and $p_{26} = 0.0321$, that are in consequence "good identifiers". Each one is present in $V_1$ and $S_2$. Besides $S_1$ is homozygote for this marker and this genotype may be hidden by $S_2$'s genotype. In consequence it is natural that $p_{12}$ and $p_{02}$ are of the same magnitude.

To compute the interesting probabilities there must be considered the following states probabilities:
- $p_{00}$: 1, 2, 3, 16, 17, 18, 19, 32, 33, 34, 35, 48, 49, 50, 51, 64, 65, 66 and 67,
- $p_{12}$: 12, 13, 14, 15, 28, 29, 30, 31, 44, 45, 46, 47, 60, 61, 62, 63, 76, 77, 78 and 79,
- $p_{10}$: 8, 9, 10, 11, 24, 25, 26, 27, 40, 41, 42, 43, 56, 57, 58, 59, 72, 73, 74 and 75,
- $p_{02}$: 4, 5, 6, 7, 20, 21, 22, 23, 36, 37, 38, 39, 52, 53, 54, 55, 68, 69, 70 and 71

from the output given by Hugin after the inserted evidence.

## 8. CRIMINAL PROBLEM DISCUSSION

Criminal identification problems are examples of situations, in which forensic approach, the DNA profiles study is usual. But the interpretation and evaluation of DNA evidence is not an easy task, see for instance, (9-10). Also, the fact that in general they are posed in probabilistic terms leads to some confusion to the judges when they have to issue a decision. In this situation the Bayesian approach is maybe the clearest to explain the significance of the evidence, see (5). And for it the use of Bayesian networks to compute the interesting probabilities is a natural option, as it was exemplified in this paper.

---
[5] It is possible to multiply the respective probabilities, for each marker, because it is assumed independence between and across marker, i.e., linkage and Hardy-Weinberg Equilibrium (8).

It is important to define which probabilities, among the possible ones to compute, interest to the problem. And in consequence to define, for each case, which hypotheses tests to implement. Of course they are Bayesian tests.

Note finally, as this example shows, that this methodology may conclude for the absolution of a suspect but not for the conviction. It only can place the suspect in the crime scene. Further work of the police must be made to conclude by the conviction or absolution.

## 9. GENERAL CONCLUSIONS

The use of networks transporting probabilities began with the geneticist Sewall Wright in the beginning of the 20$^{th}$ century (1921). In (11) it is described this new approach to problems of the kind of the one described above. The construction and use of Bayesian networks to analyse problems in forensic identification inference, was initially done there, followed by (12-14).

The civil identification problem presented obviously may occur in situations of catastrophes or accidents at which it is possible to have unidentified victims. The use of DNA evidence is quite recent in helping to solve these situations. It was shown in this work how the use of Bayesian networks is useful to evaluate that kind of evidence.

The analysis of a crime scene analogous to the considered in this work, but with two victims' and one perpetrator and two mixture traces was presented in (6, 8, 15). A problem dealing with a crime scene analogous to the one considered in this work may be seen at (16). It was also evidenced in this work how useful are the Bayesian networks in the evaluation of DNA evidence in problems of criminal identification.